# After Compilers and Operating Systems :
# The Third Advance in Application Support

Burkhard D. Burow

*Postfach 1163, 73241 Wernau, Germany*
*burow@ifh.de*
*http://www.tsia.org*

August 3, 1999

After compilers and operating systems, TSIAs are the third advance in application support. A compiler supports a high level application definition in a programming language. An operating system supports a high level interface to the resources used by an application execution. A Task System and Item Architecture (TSIA) provides an application with a transparent reliable, distributed, heterogeneous, adaptive, dynamic, real-time, interactive, parallel, secure or other execution. In addition to supporting the application execution, a TSIA also supports the application definition. This run-time support for the definition is complementary to the compile-time support of a compiler. For example, this allows a language similar to Fortran or C to deliver features promised by functional computing.

While many TSIAs exist, they previously have not been recognized as such and have served only a particular type of application. Existing TSIAs and other projects demonstrate that TSIAs are feasible for most applications.

As the next paradigm for application support, the TSIA simplifies and unifies existing computing practice and research. By solving many outstanding problems, the TSIA opens many, many new opportunities for computing.

## 1  Introduction

The Task System and Item Architecture (TSIA) is described in detail elsewhere [TSIA]. This presentation outlines the TSIA, with emphasis on its motivation and feasibility.

The compiler was the first large advance in application support. That advance earnestly began in the 1950's with the introduction of compilers for Fortran and for other programming languages. The operating system (OS) was the second large advance, earnestly beginning in the 1960's.

This presentation argues that the TSIA is the third large advance in application support. A compiler, an operating system and a TSIA each free an application from irrelevant details of computing. This motivation is described in section 2.

While many TSIAs exist, they previously have not been recognized as such and have served only a particular type of application. The TSIA approach has existed at least since the 1960's, but TSIAs earnestly began in the 1990's. Existing TSIAs and other projects demonstrate that TSIAs are feasible for most applications. This feasibility is described in section 3.

TSIAs suitable for most applications don't exist yet. By describing their motivation and feasibility, this and other initial presentations of the TSIA aim to spur the production and investigation of TSIAs in computing practice and research.

As for the compiler and the OS, the introduction of the TSIA is very much an extension, not a replacement, of existing computing practice and research. For example, part of an application may use a TSIA, while the other part does not. Similarly, few of the details of the TSIA are new. Instead most of the details are taken from elsewhere, though many are simplified or generalized.

## 2  Motivating a TSIA

### 2.1  Application Definition versus Application Execution

An application is given by its definition. For many applications, part or all of the definition is encoded in a programming language. An example fragment of such source code is `c=a+b`. An application definition also may include data or other items, but these are ignored in this presentation.

An application execution merely acts out the definition. The execution obeys the definition; the execution does not contribute to the definition. The details introduced by the execution thus are irrelevant to the definition. For example, the execution of source code requires its translation to machine code which can be executed by the computer hardware. The details of the translation, the resulting machine code and the particular computer hardware are all irrelevant to the definition. In this case, the execution details are contained in a compiler. Thus source code like `c=a+b` is free of execution details.

### 2.2  A High Level Definition

A high level definition contains only relevant details. For example, source code like `c=a+b` is relevant; the corresponding machine code is not.

A high level definition thus contains no irrelevant details. In particular, it contains few or no details of the execution. Instead, execution details are contained in a compiler, an OS, a TSIA or other systems for application support.

Of course some execution details are relevant and may be part of a high level application definition. For example, such execution details may include real-time or reliability requirements or include constraints on the time or resource costs of the execution.

The above description is illustrated in Figure 1. A high level definition only contains details relevant to the application definition. A high level definition divides the relevant from the irrelevant. Most execution details are irrelevant.

In contrast to a high level definition, a definition polluted by the irrelevant is a low level definition. As illustrated in Figure 1, such a definition contains irrelevant details in addition to the details relevant to the application definition.

### 2.3  The Division of Labour

A motivation for a high level definition may be described by the division of labour. A high level definition divides the labour required to produce the application definition from the labour required to produce the application execution. As illustrated in Figure 1, the application definition is produced by the application developer. The application execution is produced by the system developer. In addition to the initial creation, such production also includes maintenance and other efforts.

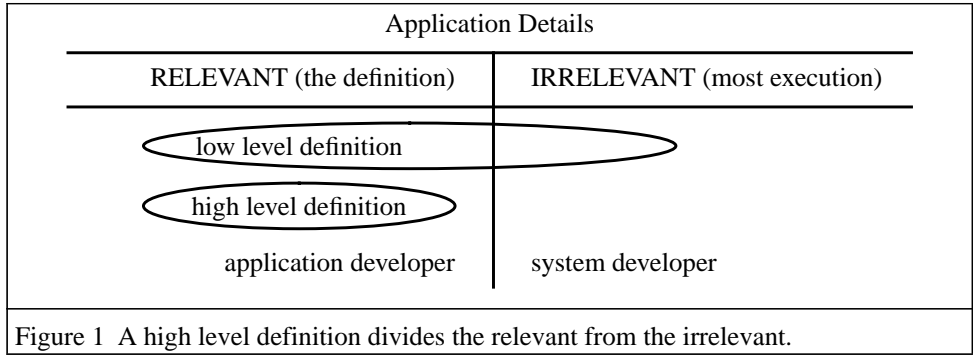

Figure 1  A high level definition divides the relevant from the irrelevant.

Since a high level definition only contains details relevant to the application definition, an application developer is concerned with the application domain. This may be media, finance, chemistry, robotics or any other use of computing.

A high level definition contains no details of the execution irrelevant to the application definition. Instead, such details are contained in a compiler, an OS, a TSIA or other system for application support. A system developer thus is concerned with the computing domain.

Dividing the application definition from the application execution greatly advances the production of an application. For example, freed from the details of one another, the production of the definition and the production of the execution are each greatly simplified. The simplification may allow for a more sophisticated and powerful application definition or execution. Similarly, the simplification allows for more qualified application developers and more qualified system developers since they do not have to be qualified as both.

Other examples of the benefits from the division of labour follow. A compiler, OS or TSIA can be contracted out, purchased off the shelf, available as free software or otherwise obtained. A compiler, OS or TSIA may or may not serve a variety of applications. If yes, then much effort can be invested in such a system since the cost is amortized against the many benefiting applications.

*2.4  Beyond the Compiler and the Operating System*

As introduced above, a compiler and an OS each contain details of the execution irrelevant to an application definition. The compiler translates a programming language to machine code. Similarly, an OS contains the details concerning the resources used by an application execution. The OS thus supports a high level interface to such use.

For some applications, a compiler and an OS suffice for a high level definition. The execution of such an application involves no execution details beyond those contained in a compiler or OS.

Other applications have an execution which requires management. Such an application has a reliable, distributed, heterogeneous, adaptive, dynamic, real-time, interactive, parallel, secure or other execution. The details of such execution management are irrelevant to an application definition and thus are not part of a high level definition.

Neither a compiler nor an OS manage an application execution. A compiler is restricted to static details. In other words, details already known at the compile-time of an application. This does not include details first known during the application execution. An OS operates resources. An OS does not operate an application execution. Thus for an application requiring execution management, a compiler and an OS suffice only for a low level definition. As illustrated in Figure 2, the definition is low level since it contains the details of execution management and these are irrelevant to the application definition.

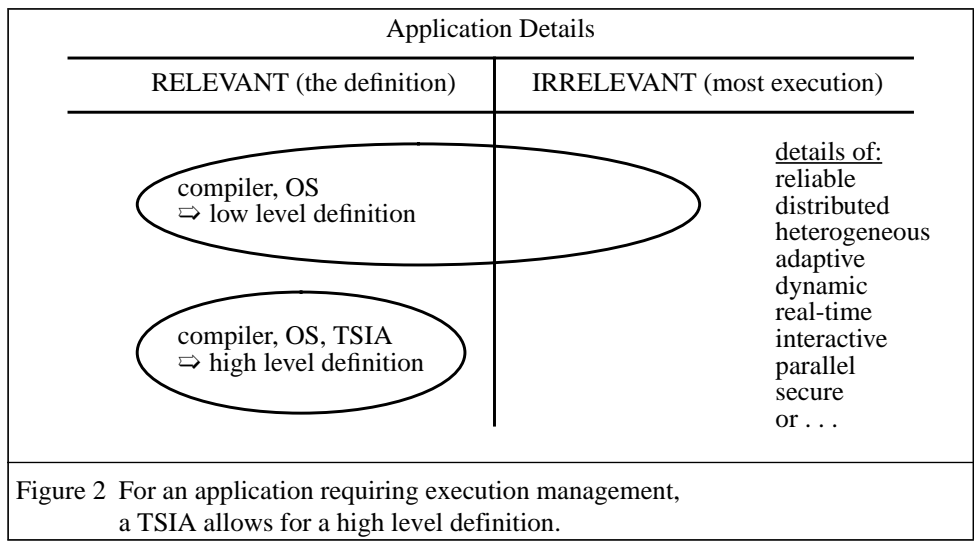

Figure 2  For an application requiring execution management,
a TSIA allows for a high level definition.

For example, a reliable execution can restart part of the execution or has redundancy. While the details are relevant to the execution, they are irrelevant to the application definition. A media, finance, chemistry, robotics or other application definition has no intrinsic interest in the details of restarting part of its execution. But with only a compiler and OS, such details of a reliable execution are part of the application definition. Since the details are irrelevant to the definition, it is a low level definition.

Similarly, an application definition has no intrinsic interest in whether it executes on 1, 2 or 300 computers in parallel. A low level definition contains such details. A high level definition is free of such details.

*2.5  A TSIA Manages Application Execution*

A TSIA manages an application execution. Thus the details of such execution management are contained in the TSIA, not in the application definition. As illustrated in Figure 2, for an application requiring such management, a compiler, OS and TSIA thus allow for a high level definition. As described above, there is great motivation for a high level definition. Since many applications require execution management, there thus is great motivation for the TSIA.

As introduced in the previous section, such execution management is required if the application has a reliable, distributed, heterogeneous, adaptive, dynamic, real-time, interactive, parallel, secure or other execution. The ability for a TSIA to manage such an execution is described in the next section.

The result of removing irrelevant execution details from an application definition is called a transparent execution. Transparency implies that the irrelevant details of the execution are not visible to the application definition.

## 3 The Feasibility of the TSIA

As described in this section, the TSIA is feasible since many TSIAs already exist. Admittedly, existing TSIAs serve only a particular type of application. However, after understanding how TSIAs work, the TSIA is generalized to most applications.

### 3.1 Many TSIAs Exist

Strong evidence for the feasibility of the TSIA is the fact that many TSIAs exist. Examples include TSIAs or TSIA-like approaches for each of the following application areas:
- a reliable, adaptive, heterogeneous, parallel application for simulation [Funnel].
- a reliable, adaptive, parallel subset of the C programming language [Cilk-NOW].
- adaptive master-worker parallelism [Linda-Piranha].
- coarse-grain parallelism [Jade].
- computational fluid dynamics on shared memory multiprocessor computers [CFD].
- using the otherwise-idle cycles of thousands of computers on the Internet to process data from the search for extraterrestrial intelligence, to crack encryption, to find primes or to execute other similar applications [Internet Computing].
- the soft instruction software architecture for real-time applications. This includes applications from the 1965 Apollo Mission and the 1970 Safeguard missile effort [SISA].

Each of the above examples achieves the all-important division between the application definition and the application execution. The first two examples are further presented in the following two subsections.

For each of the above examples, the application generally is a so-called bag-of-tasks application. This type of application is described below in subsection 3.6. Existing TSIAs generally serve this particular type of application.

TSIAs previously have not been recognized as such. This is corroborated in several ways. Firstly, since their presentations reference few predecessors, these systems seem to have been created largely independently. Secondly, the similarities between existing systems were not recognized. Thus these systems were not grouped together; though there are some exceptions [SISA]. Thirdly, systems intended to support a variety of applications have not been further pursued [Cilk-NOW][Linda-Piranha][Jade].

There seem to be at least two possible reasons why TSIAs previously were not recognized. Firstly, the achievement of a high level definition and its benefits were not recognized. Secondly, TSIAs were naively assumed to be restricted to bag-of-tasks applications. This assumption is shown below to be false, beginning in subsection 3.8.

Since TSIAs previously have not been recognized, they previously have not been pursued.

### 3.2 Funnel is a TSIA

Funnel is since 1992 the reliable, adaptive, heterogeneous, parallel batch system for the simulation application of the ZEUS experiment at DESY, the national particle physics institute in Hamburg, Germany [Funnel]. For ZEUS, Funnel currently uses the otherwise-idle cycles of about 400 workstations spread across 15 institutes around the world. Since 1995, Funnel also is the basis of the simulation system for the L3 experiment at CERN, the European particle physics institute in Geneva, Switzerland.

ZEUS and L3 are each large experiments. For example, each will last for more than ten years and each involves more than 400 physicists. To date, Funnel has served over 1000 CPU-years to ZEUS and L3. The simulation data produced by Funnel has been used in more than 100 physics publications and in many more conference presentations, theses and other analyses. Funnel thus is a production system that works well and is relied on by many people. As described in subsection 3.6, Funnel is a TSIA. The success of Funnel thus provides strong experimental evidence for the TSIA.

As a TSIA, Funnel divides the application definition from its execution. In other words, Funnel provides the simulation application with a completely transparent execution. Thus the code of the simulation application contains nothing dealing with a parallel, reliable, adaptive or heterogeneous execution. Instead, Funnel provides the simulation application with such an execution. Vice versa, Funnel contains nothing about the definition of the simulation application. Hence Funnel can execute the simulation application of ZEUS or that of L3.

Funnel is by no means the first example of a TSIA. There exist very many very similar systems [168/E][DBC][DNA][Internet Computing][Nimrod].

For the author of this presentation, his creation of Funnel began his discovery of the TSIA. His pursuit of the question "Does the success of Funnel generalize to other applications?" yields the TSIA as the answer.

### 3.3 Cilk-NOW is a TSIA

For a subset of the C programming language, Cilk-NOW provides a parallel, adaptive and reliable execution [Cilk-NOW]. Cilk-NOW is perhaps the most powerful of the existing TSIA. An aspect of this power is described in subsection 3.9.

Figure 3a) contains the code for an example application in the Cilk-NOW programming language. The example calculates the Fibonacci function. The example is taken from a Cilk-NOW presentation [Cilk-NOW].

This presentation does not describe the Cilk-NOW language. The choice of the words `thread`, `send` and `spawn` for the Cilk-NOW language is unfortunate, since their meaning in Cilk-NOW is quite different than the conventional meanings.

As a TSIA, Cilk-NOW divides the application definition from its execution. For example, Cilk-NOW provides the Fibonacci application in Figure 3a) with a completely transparent execution. Thus the code of the Fibonacci application contains nothing dealing with a parallel, reliable or adaptive execution. Instead, Cilk-NOW provides the Fibonacci application with such an execution. Vice versa, Cilk-NOW contains nothing about the Fibonacci definition nor about any other application definition. Hence Cilk-NOW can execute any application coded in the Cilk-NOW language.

In some ways the TSIA may be seen as a clean up and generalization of Cilk-NOW, including its programming language. The Cilk-NOW language is an example of a TSIA programming language. A different TSIA programming language is used in Figure 3b). That language is the TSIA language used for the remainder of this presentation. The lan-

```
a)
// Cilk-NOW language.

thread sum(cont int c,
           int a, int b)
{ send_argument(c,a+b);
}

thread fib(cont int k, int n)
{
  if (n<2)
    send_argument(k,n);
  else
  { cont int x,y;
    spawn_next sum(k,?x,?y);
    spawn fib(x,n-1);
    spawn fib(y,n-2);
  }
}
```

```
b)
// A TSIA language.

sum(int a, int b;; int c)
{
  c = a + b;
}

fib(int n;; int k)
{
  if (n<2)
    k=n;
  else
  { fib(n-1;;x);
    fib(n-2;;y);
    sum(x,y;;k);
  }
}
```

```
c)
! Fortran.

subroutine sum(a,b,c)
   integer a,b,c
   c = a + b
end

subroutine fib(n,k)
   integer n,k ,x,y
   if (n.lt.2) then
      k = n
   else
      call fib(n-1,x)
      call fib(n-2,y)
      call sum(x,y,k)
   endif
end
```

Figure 3  The same routines in the Cilk-NOW, in a TSIA and in the Fortran language.

guage is introduced in subsection 3.9 and is described in detail elsewhere [TSIA]. The language is similar to the C programming language.

The TSIA language is not yet implemented. As already mentioned in the introduction, this and other initial presentations of the TSIA aim to spur such implementations.

The TSIA code of Figure 3b) also is a Fibonacci application. It is essentially a syntactic translation of the Cilk-NOW code in Figure 3a); the meaning is largely the same. As for Cilk-NOW, a TSIA can provide the TSIA code of Figure 3b) with a transparent parallel, reliable and adaptive execution.

Except for syntax, much of Fortran is a subset of the TSIA language. This is demonstrated by translating the TSIA code of Figure 3b) into the Fortran of Figure 3c). Though similar, the Fortran code of Figure 3c) contains less information than the TSIA code of Figure 3b). This information is required by the TSIA in order to provide the definition with a transparent execution. Though not argued here, the information also benefits an application definition.

The Fortran code of Figure 3c) may be compiled and executed like any other Fortran code. The code uses recursion, which is not part of standard Fortran 77, but is supported by some compilers. For example, the SGI `f77` compiler supports recursion, while Sun's `f77 4.0` does not. A Fortran compiler does not provide the Fortran code with a transparent parallel, reliable or adaptive execution.

The above translation into Fortran demonstrates several issues. Firstly, since it obviously contains no details of its execution, the Fortran code demonstrates that TSIA code also is free of such details. Secondly, the Fortran code demonstrates that the TSIA language need not be radically different from Fortran, C or other languages. Thirdly, since it can be compiled and executed, the Fortran code demonstrates that the application definition is correct. Since the TSIA language is not yet implemented, the correctness of the TSIA code cannot yet be checked directly. Fourthly, the Fortran code demonstrates that one of the executions of TSIA code can be a conventional single computer execution.

Other executions of TSIA code, such as a parallel or adaptive or reliable execution, are described beginning in subsection 3.6.

*3.4 The Current Conventional Approach is a Low Level Definition*

The process model is the current conventional approach for an application requiring a reliable, distributed, heterogeneous, adaptive, dynamic, real-time, interactive, parallel, secure or other execution. The process model includes threads and other variations on processes. In the process model, details of the execution are an implicit part of the application definition. Since such execution details are irrelevant to an application definition, the process model yields a low level application definition. In other words, the process model does not divide the application definition from the application execution.

An example of the current conventional approach is the code of Figure 4. It is a Fibonacci application coded in the Java programming language. The code includes details for a parallel execution. The process model thus buries the Fibonacci function under many, many irrelevant execution details.

```
// Fibonacci.java : March 1997, Brandon Kearby. 1999, reformatted by B.Burow.
// http://www.eb.uah.edu/~tarique/benchmarks/Fibonacci.java

public class Fibonacci extends Thread {
  int fib;
  Fibonacci(int n) { fib = n; }
  public void run() {  // Called by start().
    if ( fib > 1 ) {
      Fibonacci thread1 = new Fibonacci(fib-1);
      Fibonacci thread2 = new Fibonacci(fib-2);
      thread1.start();   //  \__ Can execute in parallel.
      thread2.start();   //  /
      try {
        thread1.join();  //  \__ Synchronize.
        thread2.join();  //  /
        fib = thread1.getFib() + thread2.getFib();
      } catch( InterruptedException e) {
        e.printStackTrace();
      }
    }
  }
  public final int getFib() { return fib; }

  // . . .
```

Figure 4  The Fibonacci application of Figure 3, but in the Java language.

The Fibonacci application also is the example of the previous subsection. The Java definition of Figure 4 thus may be directly compared to the TSIA definition of Figure 3b). This comparison contrasts the current conventional approach against the TSIA approach advocated by this presentation. In the current conventional approach, processes and other execution details are contained within the application definition. This yields a low level definition like the Java definition of Figure 4. In the TSIA approach, processes and other execution details are contained in the support systems. This allows a high level application definition like the TSIA definition of Figure 3b). The following further compares these two paradigms for application support.

The current paradigm is the process model and is illustrated in Figure 5a). In that paradigm, the application controls its own execution. The application definition thus contains execution details. Since these execution details are irrelevant to the definition, the application definition is low level. In other words the process model does not allow for a transparent execution. The code of Figure 4 is an example of such an application. Since the application controls the execution, the execution is not controlled by the compiler, OS or other support systems. Thus these support systems can at most attempt to provide the application with a clean interface to the resources used by the application execution. This interface helps the application control the execution. For example, such interfaces simplify the starting of processes or the communication between processes.

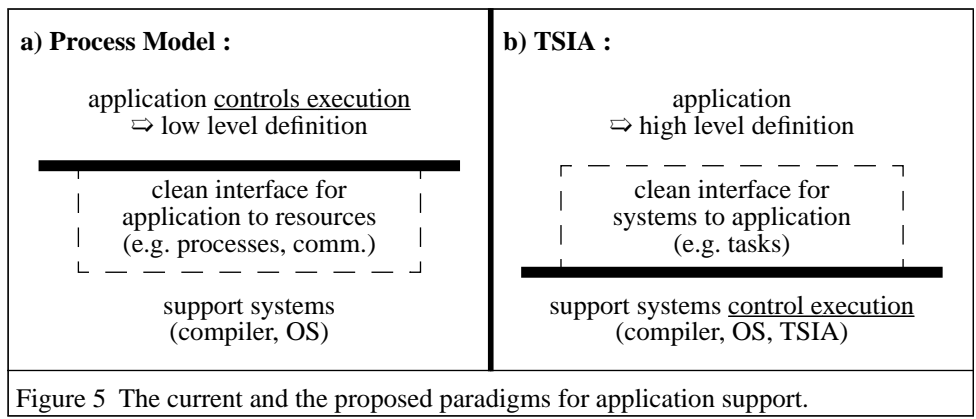

Figure 5  The current and the proposed paradigms for application support.

The paradigm proposed in this presentation is the TSIA approach and is illustrated in Figure 5b). In this paradigm, the application does not control its own execution. Since it contains no execution details, the application can have a high level definition. The code of Figure 3b) is an example of such an application.The execution is controlled by the compiler, OS, TSIA or other support systems. In order that the application definition is obeyed by the application execution, the application provides the systems with a clean interface to the application definition. As described in the next subsection, this interface is expressed in terms of tasks.

In short, in the current paradigm the application controls the execution while the support systems provide an interface. These roles are reversed in the paradigm proposed here.

*3.5  How does a TSIA work?*

As introduced in the previous subsection, in order to provide an application with a transparent execution, the TSIA requires a clean interface to the application definition. This interface requires an application to execute in terms of tasks. During its execution, a task does not communicate with other tasks.

A task consists of a computer, an instruction or routine, some arguments or data, perhaps in addition to other items. Once the items of a task are assembled, the task executes to completion. Thus the task has no control over its execution. The items of a task are declared to the TSIA. In assembling the items of the task, the TSIA has full control over the execution of the task. For an application which executes in terms of tasks, a TSIA thus controls the application execution and can provide it with a transparent execution.

Figure 6 illustrates an application execution in terms of tasks. The application is defined in terms of a programming language supported by the compiler. It in turn is supported by the item architecture (IA) part of the TSIA. The IA expresses the application definition in terms of tasks which it places into the task pool. The task system (TS) part of the TSIA takes the tasks from the pool and executes them using an OS and other resources. After execution, a task no longer exists in the task pool or elsewhere. In short, a TSIA manages the application execution. The explanation of tasks continues in the next subsection.

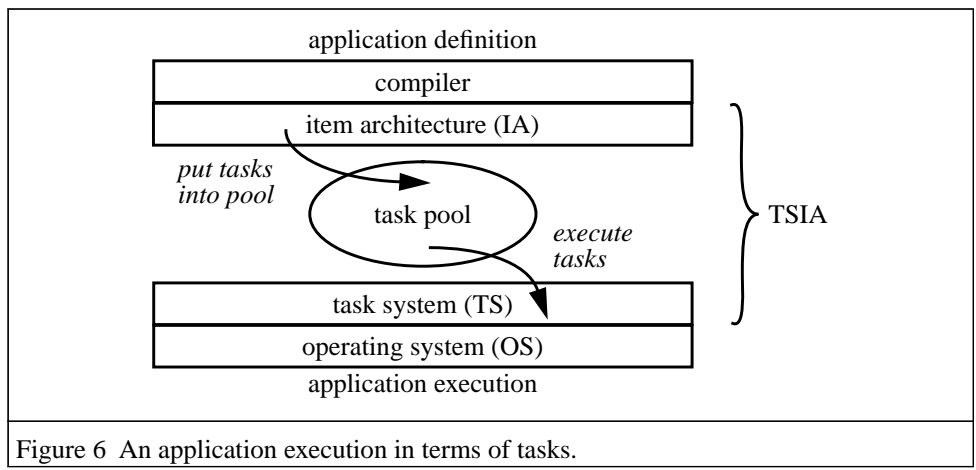

Figure 6  An application execution in terms of tasks.

Processes communicate. Tasks do not. A task thus may be considered to be a simplified process. This simplification thus may be considered to be what makes a TSIA feasible. The above situation for tasks may be compared to that for processes. A process communicates with another process in order to obtain an item. The process itself thus assembles such an item for execution. In other words, a process controls its own execution. Thus no support system can control the execution of a process. Thus for an application which executes in terms of processes, no support system can control the application execution nor provide it with a transparent execution.

*3.6 A Bag-of-Tasks Application*

In order to further explain the TSIA and its requirement that an application execute in terms of tasks, this subsection presents an application that obviously execute in terms of tasks. Such an application is called a bag-of-tasks application. Examples of such applications for which TSIAs or TSIA-like approaches exist include:
- the simulation of many independent trials or events [Funnel][Nimrod].
- the processing of many independent measurements [168/E][DBC].
- the evaluation of many independent candidate solutions [Internet Computing].
- the real-time processing of independent media frames [RTU].
- the matching of DNA to independent known sequences [DNA].

For a simple bag-of-tasks application, the input consists of independent items. Each input item is independently used to produce an independent output item. Since it is independent, the production of each item corresponds to a task. The TSIA does the simple management of each task. For example, the input of a bag-of-tasks application could be the thousand input items `in1`, `in2`, ..., `in1000`. The application then consists of a thousand independent tasks, each yielding one of the thousand independent output items `out1`, `out2`, ..., `out1000`. Before execution, these thousand tasks are in the task pool of Figure 6.

The TSIA execution of a bag-of-task application is introduced by the sequence of illustrations in Figure 7. The application simulates many independent events. A pseudocode version of the simulation program is given in Figure 7a). This application definition is used in Figure 7b) through e).

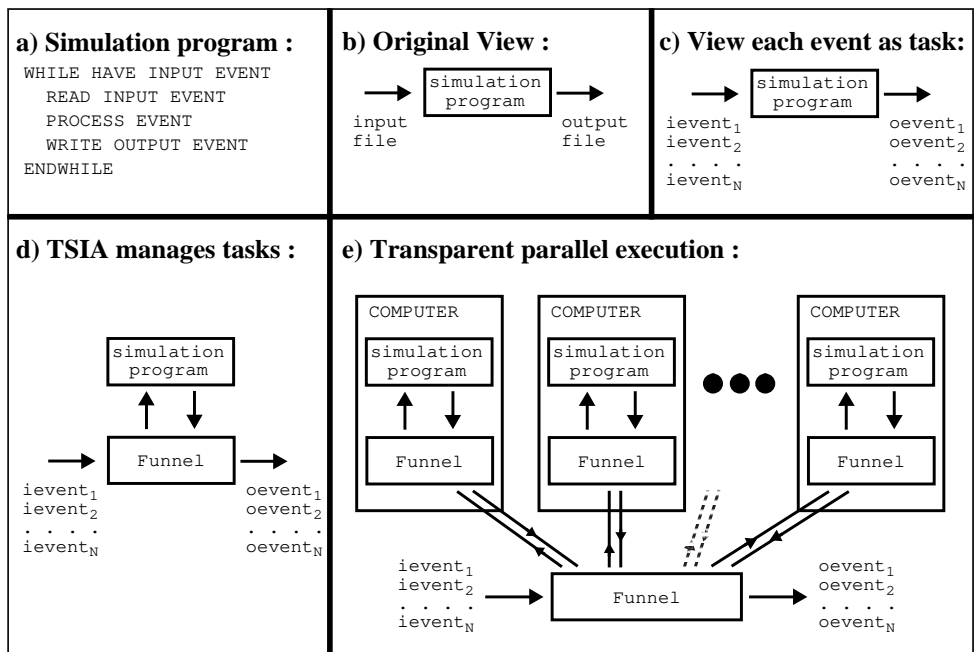

Figure 7  The TSIA execution of a bag-of-task application.

In the original view illustrated in Figure 7b), the simulation application reads an input file and writes an output file. In the TSIA approach illustrated in Figure 7c), the input file consists of independent events and each one is independently used to produce an independent output event. Since it is independent, the production of each event corresponds to a task. Thus the simulation program obviously is a bag-of-task application.

As illustrated in Figure 7d), a TSIA may be introduced to the execution in order to manage the events. In other words, the TSIA manages the tasks and thus the application execution. The TSIA here is called Funnel since this example corresponds closely to an existing TSIA by that name [Funnel].

In the execution of Figure 7d), Funnel reads an input event and passes it to the simulation program for processing. The resulting output event is returned to Funnel, which writes it out. Funnel then repeats this procedure with the next input event. The application execution is complete once all events are processed.

The execution of Figure 7d) can be a transparent reliable execution. Recall that the simulation program is defined by the pseudocode of Figure 7a). The application definition contains no details concerning reliability nor any other execution. Yet Funnel can provide the simulation application with a transparent reliable execution. For example, assume a failure of the computer on which the simulation program is executing. Funnel can restart the simulation program on the restarted computer or on another computer. Funnel then executes the remaining events. This includes the event being processed at the point of failure. For reliability, Funnel must ensure that its own execution survives failures. Like other execution details, these are within Funnel and not in the application definition.

As illustrated in Figure 7e), Funnel also can provide the simulation application with transparent parallel execution. Again recall that the simulation program of Figure 7a) contains no details concerning parallelism nor any other execution. Funnel can execute the simulation program on a number of computers. Funnel passes each input event to one of the executions of the simulation program. When an execution returns to Funnel the resulting processed output event, it receives from Funnel another input event. Like other execution details, the details of parallelism are within Funnel and not in the application definition.

In addition to this transparent parallel execution, Funnel also could provide the reliability described above. In general, a TSIA can provide a combination of executions. Other executions are described in the next section.

Funnel thus demonstrates how a TSIA divides the application definition from the application execution. The application definition is contained within the simulation program. Vice versa, the application execution is contained within Funnel.

*3.7 A Transparent Execution*

As described in the previous section, for an application which executes in terms of tasks, a TSIA can provide a transparent reliable or parallel execution. In addition to these two executions, the following examples describe some other transparent executions that can be provided by a TSIA:

- reliable execution [Cilk-NOW][DNA]. After the failure of a computer, its task executes on the restarted computer or on another computer.
- parallel execution [Cilk-NOW][DNA][Linda-Piranha]. Each task executes on one of multiple computers.
- distributed execution [DNA][Internet Computing]. A task executes on a remote computer.
- heterogeneous execution [DNA][Internet Computing]. A task executes on a different kind of computer.
- adaptive execution [Cilk-NOW][Linda-Piranha]. A reliable parallel execution allows the application execution to use a varying number of computers.
- dynamic execution [Packet Filter]. While fixed for any one task, the application definition can change between tasks.
- reactive execution [Packet Filter][RTU][SISA]. The execution of a task can meet real-time constraints.

A TSIA can provide the above transparent executions for any application which executes in terms of tasks. The feasibility is especially obvious for a bag-of-tasks application.

*3.8 Generalizing TSIAs*

As described above, a TSIA can provide a transparent reliable, distributed, heterogeneous, adaptive, dynamic, real-time, interactive, parallel, secure or other execution. Existing TSIAs generally serve bag-of-tasks applications.

Recall that a TSIA requires an application to execute in terms of tasks. During its execution, a task does not communicate with other tasks. As TSIAs are generalized to other applications, these requirements must remain.

As described in section 2, a high level definition contains only relevant details. Beyond hiding the irrelevant, a high level definition also structures relevant details using routines, arrays and other structures. Though not immediately obvious, it turns out that a TSIA can support structures and thus can serve most applications. In other words, most applications can execute in terms of tasks.

The run-time support for structures by a TSIA is complementary to the compile-time support by a compiler. The TSIA support for routines is introduced in the next section.

*3.9 Routines*

A task is made up of items: ins, inouts and outs. An in is an item required by the task. An inout is an item modified by the task. An out is an item produced by the task.

A routine can be a task. The syntax used in the TSIA language is `routine(in,..; inout,..;out,..)`. The syntax is similar to that of the Fortran or C languages, except that a semi-colon (`;`) separates the ins from the inouts and another semi-colon separates the inouts from the outs.

Obviously an out of one task can be an in of another. An example from the Fibonacci code of Figure 3b) is `fib(n-1;;x);fib(n-2;;y);sum(x,y;;k)`. As usual, the dependencies between routines are left-to-right and top-to-bottom. Thus the out `x` of `fib(n-1;;x)` is an in of `sum(x,y;;k)`, as is the out `y` of `fib(n-2;;y)`.

Not so obvious is that tasks allow for a parent-child relationship between routines. In the current conventional call of a routine, the parent communicates with the child. For example, if routine `a` calls routine `b`, then `a` will receive the outcome of `b`. This is a form of communication. Recall that during its execution, a task does not communicate with other tasks. Thus tasks do not allow for the current conventional call of a routine. Tasks thus demand an alternative implementation of the parent-child relationship.

As part of its execution, a task can replace itself by other tasks. In other words, the parent task replaces itself by the child tasks. In such a call of a routine, the parent does not communicate with the child. Tasks thus allow for a parent-child relationship between routines.

A task replacing itself by other tasks is called delegation. An example of delegation is illustrated in Figure 8. There the task pool originally contains the task `fib(10;;a)`. The code for the routine `fib(n;;k)` is given in Figure 3b). When `fib(10;;a)` is executed, it replaces itself in the task pool by the tasks `fib(9;;x)`, `fib(8;;y)` and `sum(x,y;;a)`. In other words, in its execution, `fib(10;;a)` delegates its responsibility for the out `a` to the task `sum(x,y;;a)`, which needs the outs `x` and `y` from the tasks `fib(9;;x)` and `fib(8;;y)`.

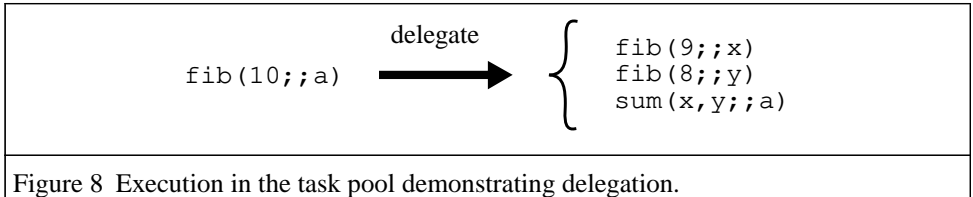

Figure 8 Execution in the task pool demonstrating delegation.

Delegation is key to much of the TSIA support for structures. Some of this support is mentioned in the next section.

Cilk-NOW is an existing TSIA which provides delegation [Cilk-NOW]. The code of Figure 3a) demonstrates delegation in Cilk-NOW. It is delegation which makes Cilk-NOW perhaps the most powerful of the existing TSIA.

Delegation is a variation on continuation, a technique from functional computing [IMPERATIVE]. A delegation may be seen as a dependence-based continuation. In contrast, the continuation of functional computing is control-based.

The execution described above for fib(10;;a) is not the only possible execution. Instead of replacing itself by other tasks, the execution of fib(10;;a) could evaluate a. In such an execution, the code of fib(n;;k) of Figure 3b) is treated like the Fortran code of Figure 3c), as described in subsection 3.3. Between these two extreme executions are many other possible executions. The variety of executions are due to the clean interface to the application definition, as provided by tasks. The TSIA is free to choose the execution which best suits the available resources and which best serves the requirements of the application.

*3.10 The TSIA For Application Definition*

The TSIA support for structures extends far beyond the support for routines described above. For example, a TSIA allows for a language similar to Fortran or C. This includes support for routines, arrays, global items and interaction. In addition, a TSIA allows such a language to deliver features promised by functional computing. This includes support for conditional items, streams, currying, ADT (abstract data type, known in TSIA as application defined type), nested routines with nonlocal items and unnamed routines. The support hides from the application definition many irrelevant details. Such detail include proper tail calling, strict or non-strict evaluation, fission and fusion, granularity, depth- or breadth-first execution, supply- or demand-driven execution, as well as a speculative or conservative execution.

The TSIA thus excellently supports the application definition. Its support for application definition thus is as strong a motivation for TSIA as is its support for application execution.

As an introduction to TSIA, this presentation can do no more than list above some of the TSIA support for an application definition. A detailed description is available elsewhere [TSIA]. Figure 9 shows an example from that description in order to present here at least some evidence for the support.

*3.11 Example: Jacobi's iterative relaxation*

This subsection approaches a real world application and includes a complete application definition. The application is taken from elsewhere and details may be found there [TSIA].

```
record Stack { push(int;;int),        // s.push(u;;e) pushes u onto the stack,
               pop(;;int,int) };      //        returning e=0 if all is well.
                                      // s.pop(;;o,e) pops o off the stack,
stack(int max;; Stack s)              //        returning e=0 if all is well.
{ int a[1:max], x=max;
  int p=0; // last full index.        // Code fragment demonstrating use.
  s.push(u;;e)                        // For simplicity, ignore error e.
     { if (p<x) { a[++p]=u; e=0; }    stack(10;;a);
       else e=1;                      stack(50;;b);
     }                                a.push(6;;e);
  s.pop(;;o,e)                        a.push(7;;e);
     { if (p>0) { o=a[p--]; e=0; }    b.push(8;;e);
       else e=1;                      a.pop(;;a1,e);   // a1 is 7;
     }                                b.pop(;;b1,e);   // b1 is 8;
}                                     a.pop(;;a2,e);   // a2 is 6;
```

Figure 9 A `Stack` ADT.

In a boundary value problem, the solution $\Phi$ must satisfy some equation in the region inside a boundary and $\Phi$ must have the given fixed values on the boundary. The solution $\Phi$, including the boundary values, may be described by an array `a`. To solve Laplace's equation, $\nabla^2\Phi = 0$, Jacobi's iterative relaxation may be used. Initially the array `a` contains the boundary values and arbitrary values for the region inside the boundary. Each iteration relaxes the array `a` towards the solution by updating each element `a(k)` inside the boundary. For a one-dimensional (1D) region and array, the relaxation is given by `a_{i+1}(k)=(a_i(k-1)+a_i(k+1))/2`, where the subscript numbers the iteration. Obviously the relaxation propagates the boundary values throughout the region. The solution given by the array `a` has converged when an iteration does not significantly change the values for the region.

Figure 10 shows a Fortran definition of Jacobi iteration for a 1D array. The definition consists principally of the two routines `jacobi` and `relax` of Figure 10b). The routine `jacobi` returns if convergence or the maximum number of iterations has been achieved. If not, `jacobi` recurses by calling the routines `relax` and `jacobi`. Using the divide-and-conquer technique, the routine `relax` divides the array into two parts, recursively calling itself on each part. The recursion ends when the array has only one element.

The application definition is presented here in Fortran for the same reasons described in subsection 3.3. The Fortran routines `jacobi` and `relax` of Figure 10b) are rewritten in the TSIA language in Figure 11. The rewriting preserves the application definition, while yielding possibilities for the execution. The definition contains no execution details. Instead, a TSIA can provide it with a transparent reliable, parallel, adaptive or other execution.

The execution of a task `jacobi` is illustrated in Figure 12. The execution of the resulting task `relax` results in two obviously independent `relax` tasks, since their inouts and outs are independent. The TSIA thus may provide Jacobi iteration with a transparent parallel execution. An execution using two computers is illustrated in Figure 13.

The execution can be efficient since it does not require any unnecessary communication. In the first iteration, the array `a` is scattered across the two computers. Subsequent iterations just 'scatter' the `relax` routine. Only the required boundary elements are communicated between iterations. The efficient execution is possible because the array `a` is

**a)**
```fortran
      program laplace
      implicit none
      integer n,imax,i
      parameter (n=8)
      real a(n),emax,e,new
      imax = 1000   ! Example
      emax = 0.0001 ! convergence cond.s.
      e = 2*emax ! Init. for jacobi().
      a(1) = 1.     ! Example
      a(n) = n      ! boundary cond.s.
      do i = 2,n-1  !
        a(i) = 0.   ! Define array.
      enddo         !

      call jacobi(n,emax,e,imax,a)

      print *,e,' was highest change.'
      print *,imax,' iterations remained.'
C Check jacobi() by checking convergence.
      do i = 2,n-1
        new = (a(i-1)+a(i+1))/2
        if (abs(new-a(i)) .gt. emax) then
          print *,'a(',i,') no converge.'
        endif
      enddo
      end
```

**b)**
```fortran
C a(1) and a(n) are fixed.
C Relax a(2:n-1) for imax iterations
C or until convergence:
C         abs(new-old)<emax for each a(i).
C Relax means a(i) = (a(i-1)+a(i+1))/2.
C Initially requires e>emax.

      subroutine jacobi(n,emax,e,imax,a)
      implicit none
      integer n,imax
      real a(n),e,emax
C Convergence or max iterations?
      if (e .lt. emax) return
      imax = imax - 1
      if (imax .lt. 0) return
C Otherwise another relaxation iteration.
      call relax(n-2,a(1),a(n),a(2),e)
      call jacobi(n,emax,e,imax,a)
      end

      subroutine relax(n,m,p,a,e)
      implicit none
      integer n,k
      real a(n),m,p,e,mk,pk,olda,em,ep
      if (n .eq. 1) then
        call set(a,olda)
        call avg(m,p,a)
        call absdiff(olda,a,e)
      else
        k = n/2
        call set(a(k  ),mk)
        call set(a(k+1),pk)
        call relax(k  ,m ,pk,a     ,em)
        call relax(n-k,mk,p ,a(k+1),ep)
        call maxi(em,ep,e)
      endif
      end
```

**c)**
```fortran
      subroutine set(a,b)
      real a,b
      b = a
      end

      subroutine maxi(a,b,c)
      real a,b,c
      if (a .gt. b) then
        c = a
      else
        c = b
      endif
      end

      subroutine avg(a,b,c)
      real a,b,c
      c = (a + b)/2.
      end

      subroutine absdiff(a,b,c)
      real a,b,c
      c = abs(b-a)
      end
```

Figure 10 Fortran application for Jacobi iteration in 1D.

```
set(    real a         ;; real b);              jacobi(int n, real emax;
maxi(   real a, real b;; real c);                       real e, int imax, del real a[n];) {
avg(    real a, real b;; real c);               // Convergence or max iterations?
absdiff(real a, real b;; real c);               if (e < emax) return;
                                                imax = imax - 1;
relax(int n, del real m, del real p;            if (imax < 0) return;
      del real a[n]; del real e) {              //Otherwise another relaxation iteration.
if (n == 1) {                                   relax(n,a[1],a[n];a[2:n-1];e);
  set(a;;olda);                                 jacobi(n,emax;e,imax,a;);
  avg(m,p;;a);                                  }
  absdiff(olda,a;;e);
{
  int k = n/2;
  set(a(k  );;mk);
  set(a(k+1);;pk);
  relax(k  ,m ,pk;a      ;em);
  relax(n-k,mk,p ;a(k+1);ep);
  maxi(em,ep;;e);
}
}
```

Figure 11 TSIA routines replacing the Fortran routines of Figure 10b).

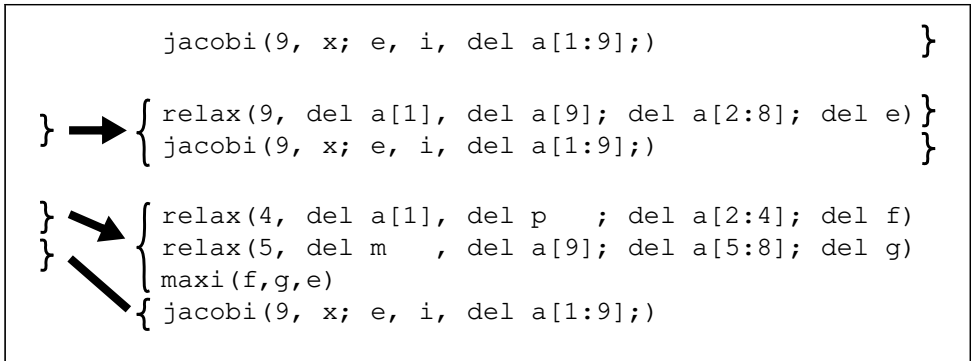

Figure 12 Execution in the task pool of Jacobi iteration of Figure 11.

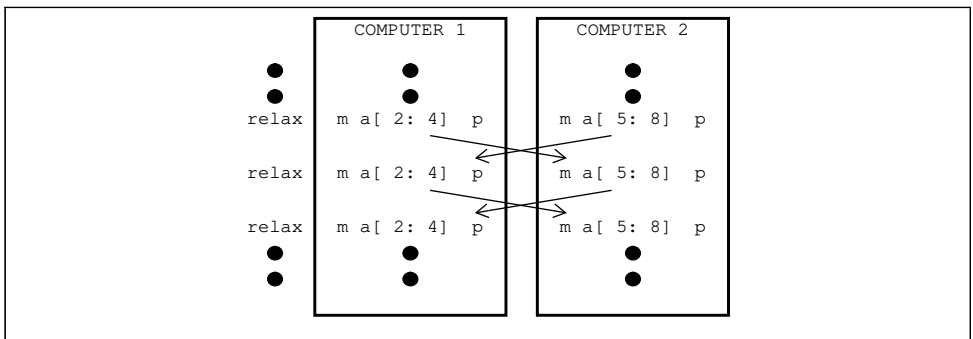

Figure 13 A parallel execution of Jacobi iteration.

delegated by the routine `jacobi`. The same holds for the routine `relax`. The delegation is declared to the TSIA by the keyword `del`. Since it delegates the array `a`, the routine `jacobi` executes without accessing the array `a`. Thus the array `a` need not be gathered for the execution of the routine `jacobi`. Thus the array `a` need not be unnecessarily gathered and scattered between iterations.

In addition to a parallel execution, the TSIA can transparently provide Jacobi iteration with a reliable, adaptive or other execution.

## 4 Conclusion

A motivation for TSIA is its support for a high level application definition. As described in section 2, a TSIA provides an application with a transparent reliable, distributed, heterogeneous, adaptive, dynamic, real-time, interactive, parallel, secure or other execution. As required by a high level definition, the irrelevant details of the execution thus are not contained in the application definition. As described in section 3, a TSIA also supports the structures required for a high level definition.

Also described in section 3 is the feasibility of the TSIA. The many existing TSIAs demonstrate that a TSIA is feasible for an application which executes in terms of tasks. During its execution, a task does not communicate with other tasks. Since the structures of a high level application definition can execute in terms of tasks, the TSIA is feasible for most applications.

The TSIA simplifies and unifies computing practice and research. An example is the simplicity gained by separating the application definition from the application execution. Two examples of unification follow. The TSIA unifies reliable, distributed, heterogeneous, adaptive, dynamic, real-time, interactive, parallel, secure and other executions. The TSIA unifies imperative and functional computing, thus allowing a language similar to Fortran or C to deliver features promised by functional computing.

By solving many outstanding problems, the TSIA opens many, many new opportunities for computing. Is a TSIA able to deliver objects, persistence, polymorphism and other features not yet examined for TSIA? Will TSIA lead to new structures and other new features for defining an application? How can an application best define real-time, reliability, cost or other requirements on the execution?

Because it presently requires a great deal of effort, few existing applications have a reliable, distributed, heterogeneous, adaptive, dynamic, real-time, interactive, parallel, secure and other execution. By transparently providing such an execution, what new applications are made possible by a TSIA? Similarly, when the use of any resource becomes transparent, what resources will be used and what new resources will become available?

Just as the implementation of compilers and OSs has provided many challenges and successes, so will the implementation of TSIAs. Compilers and OSs are indispensable parts of present-day computing. What will computing be like when TSIAs also are an indispensable part of computing?

**Acknowledgments**

This presentation is based on invited seminars held in May-July 1999 at the IBM Development Lab in Böblingen, the Lawrence Livermore National Lab. (LLNL), the Center for Advanced Computing Research (CACR) of the California Institute of Technology (Caltech), the Stanford Linear Accelerator Center (SLAC), the Fermi National Accelerator Laboratory (Fermilab), the Argonne National Laboratory, the Deutsches Elektronen-Synchrotron (DESY), as well as at the European Laboratory for Particle Physics (CERN).